# Imaging the dynamics of individual hydrogen atom intercalated between two graphene sheets

Wen-Xiao Wang[1,§], Yi-Wen Wei[2,§], Si-Yu Li[1], Xinqi Li[3], Xiaosong Wu[3], Ji Feng[2,*] and Lin He[1,*]

[1]Center for Advanced Quantum Studies, Department of Physics, Beijing Normal University, Beijing, 100875, People's Republic of China

[2]International Center for Quantum Materials, School of Physics, Peking University, Beijing 100871, P.R.China

[3]State Key Laboratory for Artificial Microstructure and Mesoscopic Physics, Peking University, Beijing 100871, China and Collaborative Innovation Centre of Quantum Matter, Beijing 100871, China

[§]These authors contributed equally to this work.

*Correspondence and requests for materials should be addressed to J.F. (e-mail: jfeng11@pku.edu.cn) and L.H. (e-mail: helin@bnu.edu.cn).

**The interlayer gallery between two adjacent sheets of van der Waals materials is expected to modify properties of atoms and molecules confined at the atomic interfaces. Here, we directly image individual hydrogen atom intercalated between two graphene sheets and investigate its dynamics by scanning tunnelling microscope (STM). The intercalated hydrogen atom is found to be remarkably different from atomic hydrogen chemisorbed on external surface of graphene. Our STM measurements, complemented by first-principles calculations, show that the hydrogen atom intercalated between two graphene sheets has dramatically reduced potential barriers for elementary migration steps. Especially, the confined atomic hydrogen dissociation energy from graphene is reduced to 0.34 eV, which is only about a third of a hydrogen atom chemisorbed on graphene. This offers a unique platform for direct imaging of the atomic dynamics of confined atoms. Our results suggest that the atomic interfaces of van der Waals materials may provide a confined environment to tune the interfacial chemical reactions.**

Two-dimensional (2D) materials, such as graphene, expose all the atoms to the surfaces, therefore, adsorbed atoms or molecules can dramatically change their atomic and electronic structures[1-7]. For example, hydrogen atoms adsorbed on graphene can generate energy gaps[1,2] and even induce magnetic moments[4-6] in graphene. Although the strong influence of adsorbates on graphene has attracted tremendous interest, the effects of graphene on the adsorbed atoms or molecules are much less explored. Very recently, several groups began to study effect of interlayer gallery between two graphene sheets on entrapped atoms and molecules, and demonstrated that the confined space can modify their structures and properties[8-10]. However, probing the dynamics and chemical reaction of atoms intercalated in van der Waals materials proves extremely challenging. Here we directly image an individual hydrogen atom intercalated between two graphene sheets and study its dynamics by using scanning tunnelling microscope (STM). Our result demonstrates that the hydrogen atom confined at the atomic interface behaves remarkably different from that chemisorbed on the surface of graphene. For the former case, the hydrogen atom between two graphene sheets is found to be extraordinarily mobile at 78 K, which can be further enhanced by the STM tip, to allow for observation of migratory dynamics, whereas the chemisorbed hydrogen on the surface of graphene is localized at such a low temperature. Our first principles calculations indicate that potential barriers for hydrogen dynamics, especially the barriers for desorption, are much reduced when the hydrogen atom is confined between two graphene sheets.

In this work, the bilayer graphene was grown on the SiC by thermal decomposition with the hydrogen assisted growth method. The hydrogen, as a carbon etchant, not only suppresses the growth of multilayer graphene, but also suppresses the defect formation and nucleation of graphene, which results in smooth surface morphology of graphene[11]. We chose to study hydrogen atoms confined between two graphene sheets based on following reasons. (I) The hydrogen atoms exist naturally in the sample synthesized by our method (they are generated in the dissociation process of both $CH_4$ and $H_2$). (II) It is convenient to identify the chemisorbed hydrogen atoms, the simplest adsorbate species, on graphene by using STM. (III) The chemisorption and desorption of individual hydrogen atoms on graphene, to some extents, maybe the simplest chemical reaction confined between two graphene sheets, which can be treated as a model system to study the interfacial reactions.

**Results**

**The structure of individual hydrogen atom intercalated between two graphene sheets.**

The hydrogen atom intercalated between two graphene sheets has multiple local potential energy minima. It could be attached to either the first or the second layer, as illustrated in Fig. 1a and 1d (side views). The obtained STM images for the two cases are quite different. In the former case, the high-resolution STM image (Fig. 1b) exhibits threefold symmetry and a $\sqrt{3} \times \sqrt{3} R 30°$ (R3) interference pattern, which are well reproduced in our simulated STM image (Fig. 1c). The major features of our experimental result, such as a bright triangle center and the interference pattern around the chemisorbed H, are similar to those reported for H chemisorbed on external surface of graphene in previous studies[4,6]. In the latter case (Fig. 1d), the intercalated hydrogen atom in the STM image becomes an atomic-sized bright dot, as shown in Fig. 1e. Such an experimental feature and the bias dependent STM images, as shown in Fig. 2, are also reproduced quite well by the corresponding simulated images shown in Fig. 1f and in Fig. 2d-2f. Therefore, isolated hydrogen atom confined between bilayer graphene exhibits different features when chemisorbed on the first and the second layer.

We can exclude atomic defects, substitutional atoms, and other types of intercalation species as the origin of the experimental features in Fig. 1. A single carbon vacancy can generate similar features in graphene except that the symmetry is broken due to the Jahn-Teller distortion[12] (see Supplementary Fig. 1). This slight difference could help us to distinguish them in the STM measurements. The substitutional atoms, for example the N-substitution[13,14], are very stable and localized, which distinct from our experimental observation and allow us to completely rule them out. The other possible intercalation species, such as the O, Si, Ne atoms, exhibit quite different features, like a bright 'blister' and without apparent scattering pattern[15-17], in the STM measurements comparing to that of H chemisorbed on graphene. We can also rule out the hydrogen atom chemisorbed on external surface of the first layer as the origin of the observed features in Fig. 1 based on the following three experimental results. First, the hydrogen atom on surface of graphene would have a higher height profile, > 250 pm, than the topographically small features observed, < 200 pm (see Fig. 2 and supplementary Fig. 2 for details). Second, the hydrogen atom on external surface of graphene would possibly get swept away by the STM tip during the measurements, however, this does not happen during our STM measurements for more than 30 hours. Third, in our experiment, we frequently observed transition of the hydrogen atom between chemisorbed on the first layer and the second layer, i.e., the recorded STM images of the adatom changes between that of Fig. 1d and that of Fig. 1e. These results demonstrated explicitly that the hydrogen atom intercalated between the two graphene sheets is the origin of the observed features in our STM measurements.

**Dynamics of individual hydrogen atom intercalated between two graphene sheets.**

Unlike single carbon vacancies, which are difficult to generate in graphene (the formation energy is as high as about 7.4 eV)[12,18,19], the hydrogen atoms are very easy to chemisorb on graphene (the surface adsorption barrier on graphene is only about 0.2 eV)[1,2,5,6,20-22]. However, to break the C-H bond for chemisorbed hydrogen dissociation from graphene, one has to overcome a relatively higher potential barrier of about 1.1 eV[23,24]. Therefore, the chemisorbed hydrogen atoms on the external surface of graphene are quite stable at low temperature. At 78 K (the temperature at which our experiments were performed), we find that they can be stable for more than 10 hours during the STM measurements. This stabilization allows the patterning and characterization of the chemisorbed hydrogen atoms on external surface of graphene by using STM in previous studies[6,22].

For the hydrogen atom confined between two graphene sheets, however, our experiment indicates that it is quite easy to observe its desorption and migration. Figure 3 shows a representative diffusion process of the H atom (highlighted by yellow circle) chemisorbed on the second graphene layer, in which the arrows roughly denote the moving directions. Obviously, the H atom changes its positions continuously during the STM measurement. For the structure shown in Fig. 3, there are two advantages to study the dynamics of the single hydrogen atom. First, there are two stable and slightly separated bright protrusions, which may arise from defects of the substrate. The origin of the coordinate and the x, y axes are defined based on the two bright protrusions, as shown in Fig. 3a. This enables us to explicitly define the relatively positions of the hydrogen atom. Second, there is only one hydrogen atom in the studied region ~ 50 nm × 50 nm, which enables us to trace the dynamics of the hydrogen atom. In our STM measurements with low bias voltages, we observed the diffusion processes occasionally and randomly. However, by using a bias voltage or bias pulse larger than 1 V, we can stimulate the hydrogen atom locked between the two graphene sheets and observe its dynamics in a more controlled way (see Fig. 3 and Supplementary Movies). By using this method, we can even capture the moving path of the H atom during scanning at bias of 1 V (see Fig. 3c and Supplementary Fig. 3).

Besides the diffusion of the confined H atom, we also frequently observed transition of the hydrogen atom between chemisorbed on the first layer and the second layer in our experiment. Figure 4 shows several typical STM images recorded consecutively in our experiment. In Fig. 4a, the hydrogen atom chemisorbed on the inner surface of the first layer. It chemisorbed on the second layer in Fig. 4c. Interestingly, we observe the in-between state in Fig. 4b, where the hydrogen atom desorbed from the first layer and then chemisorbed on the second layer. Such a process is also explicitly shown in Fig. 4f by the profile lines across the hydrogen atom. It is very interesting to note that all the R3 scatter patterns generated by the H atom chemisorbed on the first layer are in the same direction in our experiment (see Fig. 4a and 4d for examples). This indicates that the H prefers to adsorb one kind of sublattice of the first graphene sheet because of that the directions of R3 scatter patterns for H chemisorbed on two sublattices of graphene are different, as shown in Supplementary Figure 6. Such a feature is quite different from that of hydrogen atom chemisorbed on the external surface of graphene, where the probabilities for the two sublattices should be equal. We will demonstrate subsequently that this feature is a direct experimental result of the H atom intercalated between two graphene sheets.

**Theoretical studies of the dynamics of the H atom intercalated between two graphene sheets.**

The above result clearly indicates that the potential barriers of diffusion and desorption are much reduced for the confined hydrogen atom. To validate this hypothesis, we calculated the diffusion and desorption processes of a hydrogen atom (a) chemisorbed on external surface of graphene and (b) intercalated between two graphene sheets (Figure 5). In the calculations, the two graphene sheets are assumed to be Bernal stacked and free standing. From the calculation, we can see that the barrier of diffusion for a hydrogen atom chemisorbed on external surface of graphene is about 1.02 eV (Supplementary Figure 4), whereas the barrier of diffusion for a confined hydrogen atom is only 0.75 eV, as shown in Fig. 5b. Our calculation further indicates that an effective electric field on the sample may further reduce the barrier (Supplementary Figure 5). In the experiment, both the STM tip and substrate could generate an effective electric field on the sample[25-27]. In our STM measurements, we observe the dynamics of the hydrogen atom more frequently by using a bias voltage larger than 1 V, which may partially arise from the lowering of the potential barrier induced by the effective electric field. Obviously, these barrier reductions could promote the diffusion of the hydrogen atom locked between the two graphene sheets.

For a hydrogen atom chemisorbed on external surface of graphene, the potential barrier of desorption is about 1.1 eV[23,24]. Our calculation indicates that the potential barrier is reduced if the hydrogen atom is locked between two graphene sheets. There are two possible and basic routes for the desorption process of the confined hydrogen atom, as shown in Fig. 5c and Fig. 5d. In the first case that the hydrogen atom desorbs from one sublattice of the first layer, which lies above the centre of a hexagon in the second layer, and attaches itself to the closest carbon in the second layer, the potential barrier is only reduced to about 0.91 eV (Fig. 5c). However, in the second case that the hydrogen atom desorbs from the sublattice of the first layer, which lies above the atoms in the second layer, and then chemisorbs its nearest-neighbour in the second layer, the potential barrier is only 0.34 eV (Fig. 5d). The remarkable reduction of barrier is expected to drastically increase desorption frequency of the confined hydrogen atom, which ensures the direct imaging of the atomic dynamics in our STM measurements. The potential barrier for the second case is much lower than that of the first case, indicating that the desorption and chemisorption processes prefer to occur on the sublattice above the C atoms in the adjacent graphene sheet, i.e., prefer to occur on one kind of sublattice of graphene. This is in good agreement with our STM observations shown in Fig. 4 and provides further evidence that the H atom is confined between two graphene sheets.

**Discussion.**

In our experiment, we systematically studied dynamics of the confined hydrogen atom for more than 30 hours and observed dozens of times of its desorption and diffusion between the two graphene sheets. However, the confined hydrogen atom never escapes from the intercalation. Our result provides an atomic-level evidence that graphene is an impermeable atomic membrane at 78 K, even for the lowest-atomic-number H atom[28-33]. Very recently, Hu, et al. shows that graphene monolayer is permeable to thermal protons under ambient conditions[34]. They find that the proton-penetration conductivity at 330 K is very high, whereas it almost reduces to zero when the temperature decreases to 270 K. According to their study, the relation of proton-penetration conductivities with the temperature follows Arrhenius-type behavior, $e^{-E/k_BT}$. Obviously, thermal fluctuation plays an important role in the penetration. It is also important to emphasize that the barrier of the penetration for physisorption proton is only 1.41 eV (the case in Ref. 34), however, the potential barrier increases to about 4.54 eV once a C–H bond is formed[28] (our case). Therefore, the hydrogen atom cannot penetrate the graphene monolayer in our experiment.

In summary, we directly imaged a hydrogen atom intercalated between two graphene sheets and studied its dynamics systematically. Our STM experiments, complemented by first-principles calculations, indicate that the potential barriers of diffusion and desorption for the confined hydrogen atom are reduced unexpectedly. Such a result suggests that the atomic interfaces of graphene multilayers and other van der Waals materials may provide a confined environment to tune the interfacial reactions.

## Methods
### Sample preparation and STM measurements.

We prepared a high-quality graphene film on a silicon carbide crystal by thermal decomposition process. On-axis semi-insulating 4H-SiC ($000\bar{1}$) wafers were purchased from Cree Inc, and the mis-cut angle was estimated to be 5′. Prior to growth, SiC chips were hydrogen etched at 1600 ℃ for 20 min to remove polishing scratches, so we obtained atomically flat SiC substrate surface. The growth was carried out in a home-made high vacuum induction furnace. Samples were first annealed in vacuum at about 1000 ℃ for 60 min to remove the native oxide on the surface. Then growth took place in a mixed gas flow (2% hydrogen and 98% argon) at 20-40 sccm. The hydrogen partial pressure is about 0.03 mbar. After 15 min of growth at about 1500 ℃, heating was shut off and the samples were allowed to cool naturally.

The scanning tunneling microscopy (STM) system is an ultrahigh vacuum single-probe scanning probe microscope (USM-1500) from UNISOKU. All STM measurements were performed at liquid-nitrogen temperature and the images were taken in a constant-current scanning mode. The STM tips were obtained by chemical etching from a wire of Pt(80%) Ir(20%) alloys. Lateral dimensions observed in the STM images were calibrated using a standard graphene lattice as well as a Si (111)-(7×7) lattice.

### Calculation.
The spin-polarized density functional theory (DFT) calculations were performed with the Vienna ab initio simulation package(VASP)[35,36], using the projector augmented wave (PAW) potentials[37,38]. The electron-electron exchange correlation was approximated by the generalized gradient functional of Perdew-Burke-Ernzerhof (PBE)[39]. We used a vacuum space of 1.5 nm along the z direction. A cutoff energy of 400 eV was used together with a Gamma centered 8×8×1 k-points grid[40] for a (4×4) supercell containing 64 carbon atoms of bilayer graphene. The van der Waals interactions were described via the DFT-D2 method of Grimme[41]. Energy barriers were determined using the climbing-image nudged elastic band (CI-NEB) method[42,43]. STM images were simulated using the Tersoff-Hamann formalism with a (8×8) supercell (256 carbon atoms) of bilayer graphene.

## Acknowledgements

This work was supported by the National Natural Science Foundation of China (Grant Nos. 11674029, 11422430, 11374035, 11334006, 11222436), the National Basic Research Program of China (Grants Nos. 2014CB920903, 2013CBA01603), the program for New Century Excellent Talents in University of the Ministry of Education of China (Grant No. NCET-13-0054). L.H. also acknowledges support from the National Program for Support of Top-notch Young Professionals.

## Author contributions

W.X.W. and S.Y.L. performed the STM experiments and analyzed the data. X.Q.L. and X.S.W. synthesized the sample. Y.W.W. performed the theoretical calculations. L.H. conceived and provided advice on the experiment, and analysis. J.F. provided

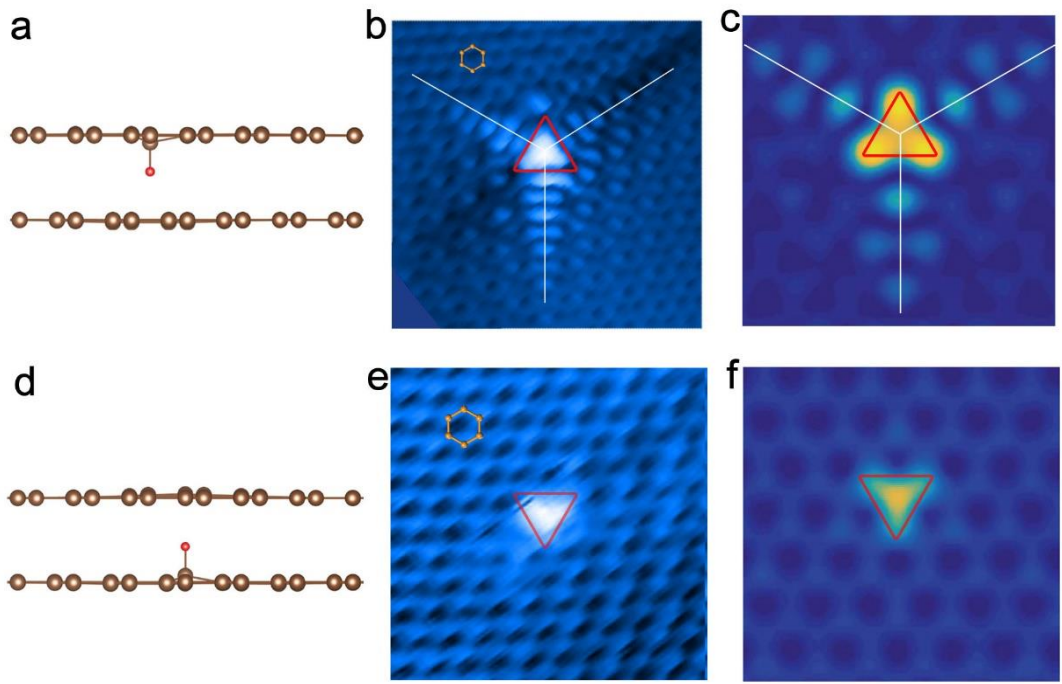

**Figure 1 | The structure of H confined between graphene layers. (a,d)** The side view for ball-and-stick model of bilayer graphene showing the position of the H atom adsorbing. The red ball denotes the hydrogen atom. **(b,e)** Typical atomic STM images of hydrogen atom chemisorbed on epitaxial graphene on $SiC(000\bar{1})$, corresponding to schematic diagrams shown in **a** and **d** respectively. Sample Bias and set points are (0.4 V, 300 pA) and (0.07 V, 300 pA) respectively. The honeycomb structures of graphene are overlaid onto the STM images. **(c,f)** The simulated STM images at comparable imaging conditions show the similar features to that in **b** and **e**, respectively.

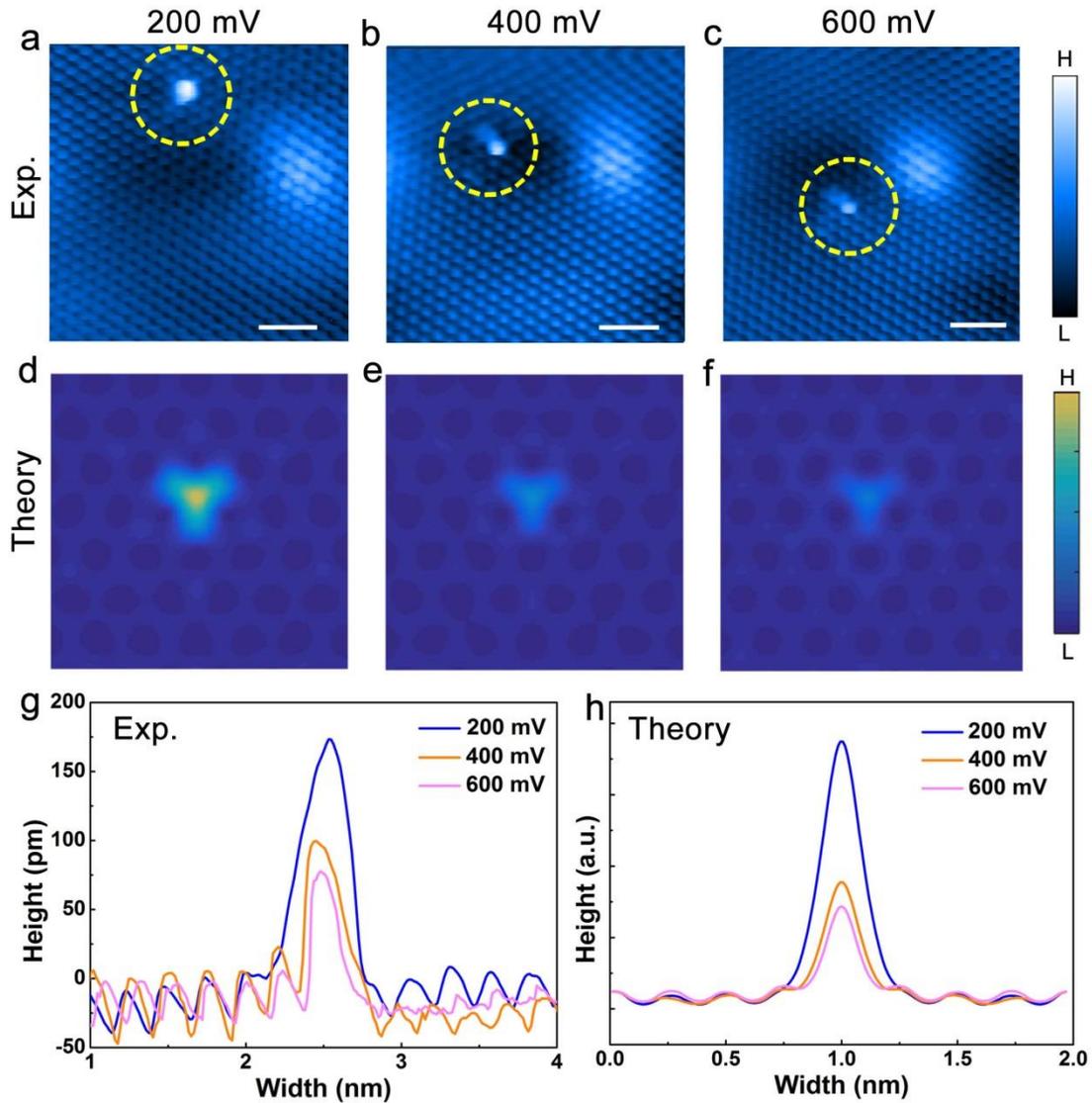

**Figure 2 | The height of H intercalated between graphene bilayer. (a-c)** Atomic STM images of hydrogen atom (highlighted by yellow circle) chemisorbed on the second layer of graphene. The images are recorded at sample bias of 200 mV, 400 mV and 600 mV, respectively. The larger and stable bright protrusion may arise from defect of the SiC substrate. **(d-f)** The corresponding simulated images at comparable imaging conditions to that of **a-c**. They are adjusted into the same contrast for better comparison. **(g,h)** show the profile lines across the hydrogen atom of **a-c** and **d-f,** respectively.

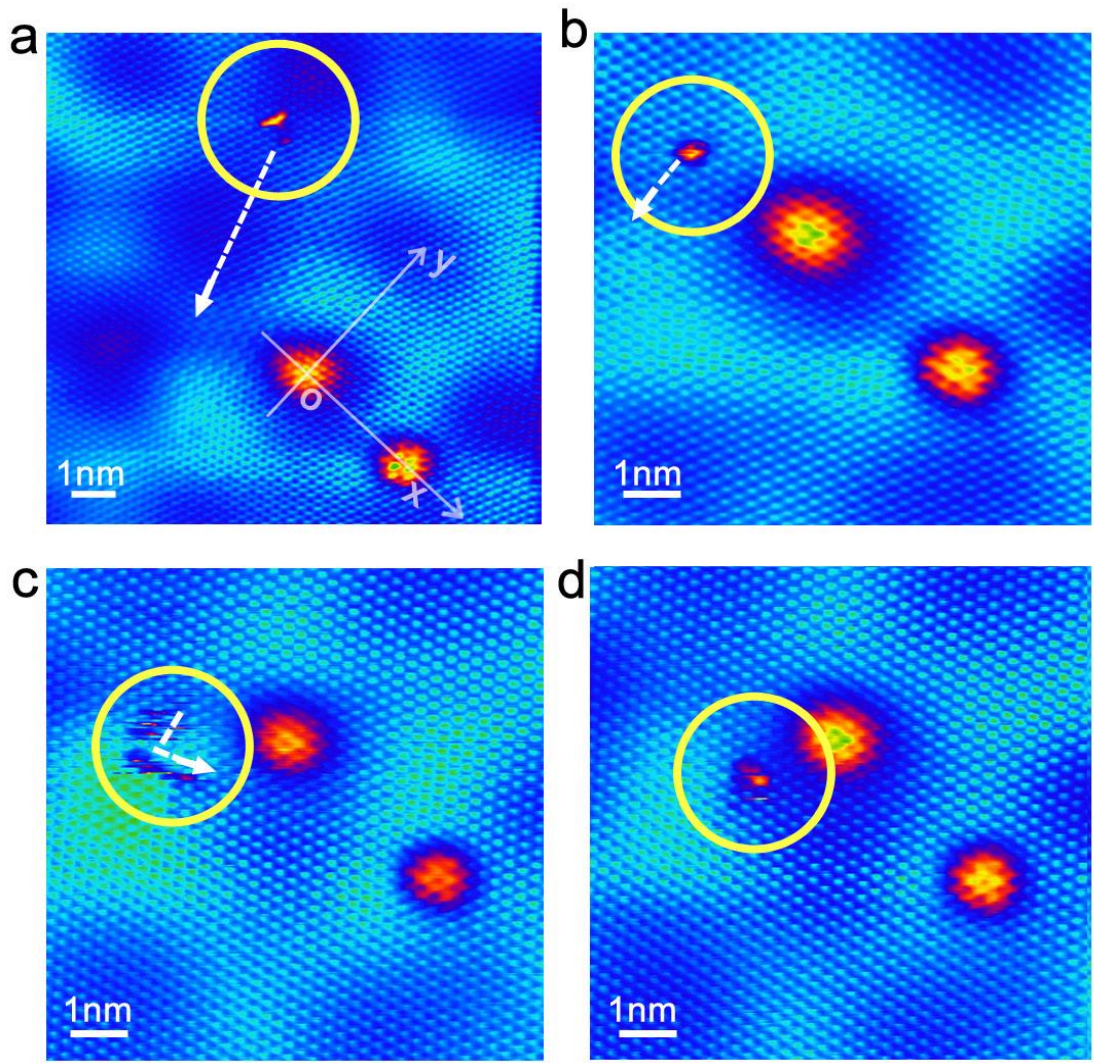

**Figure 3 | The diffusion of the confined hydrogen atom. a-d** The series of STM images showing the diffusion process of the confined hydrogen atom (highlighted by yellow circle) at 78 K. The line across two large protrusions (defects of the substrate) is defined as x axis and the perpendicular direction is the y axis. Sample Bias and setpoints for **a**, **b**, and **d** are (0.6 V, 300 pA); **c** (1 V, 300 pA).

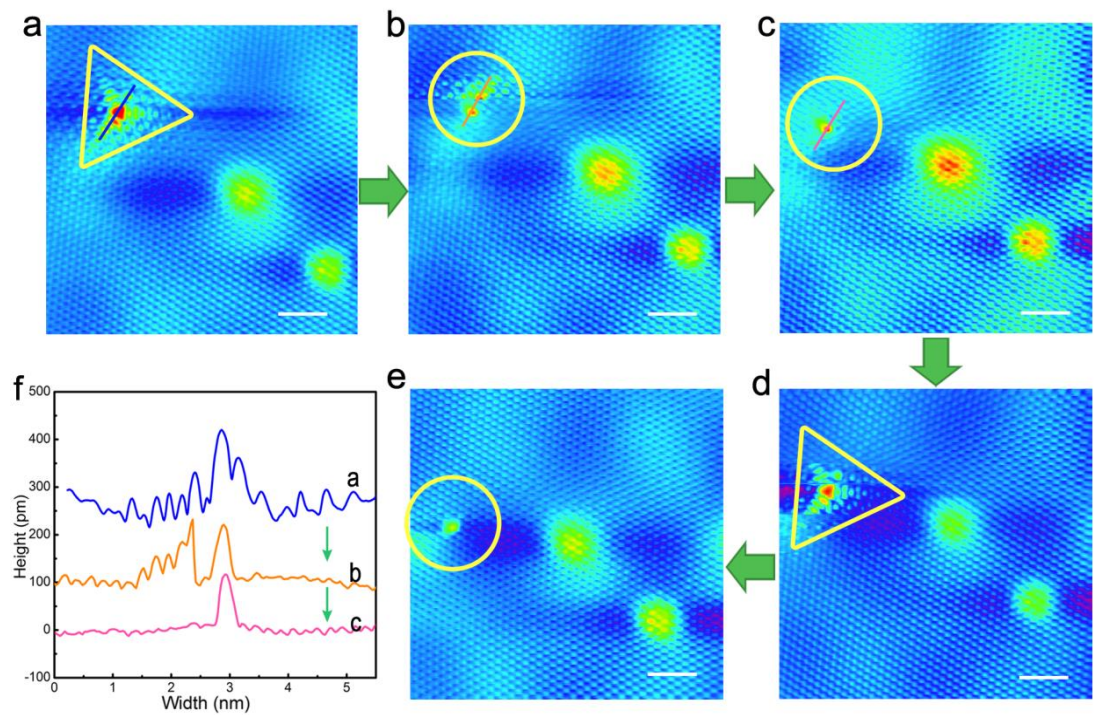

**Figure 4 | The desorption and chemisorption processes of the confined H atom. (a)** The hydrogen atom chemisorbed on the inner surface of the first layer. **(b)** The in-between state of the hydrogen atom, which changes from that chemisorbed on the inner surface of the first layer to that chemisorbed on the second layer. **(c)** The hydrogen atom chemisorbed on the second layer. **(d,e)** The hydrogen atom chemisorbed on the inner surface of the first layer and chemisorbed on the second layer respectively. Sample Bias and setpoint are (0.4 V, 300 pA) for **a-d**, and ( 0.6 V, 300 pA) for **e**. Scale bar is 2 nm. We use triangle and circle to denote that hydrogen atom chemisorbed on the inner surface of the first layer and chemisorbed on the second layer respectively. **(f)** Line profiles across the H atom taken from **a**, **b** and **c** to illustrate the topographic changes.

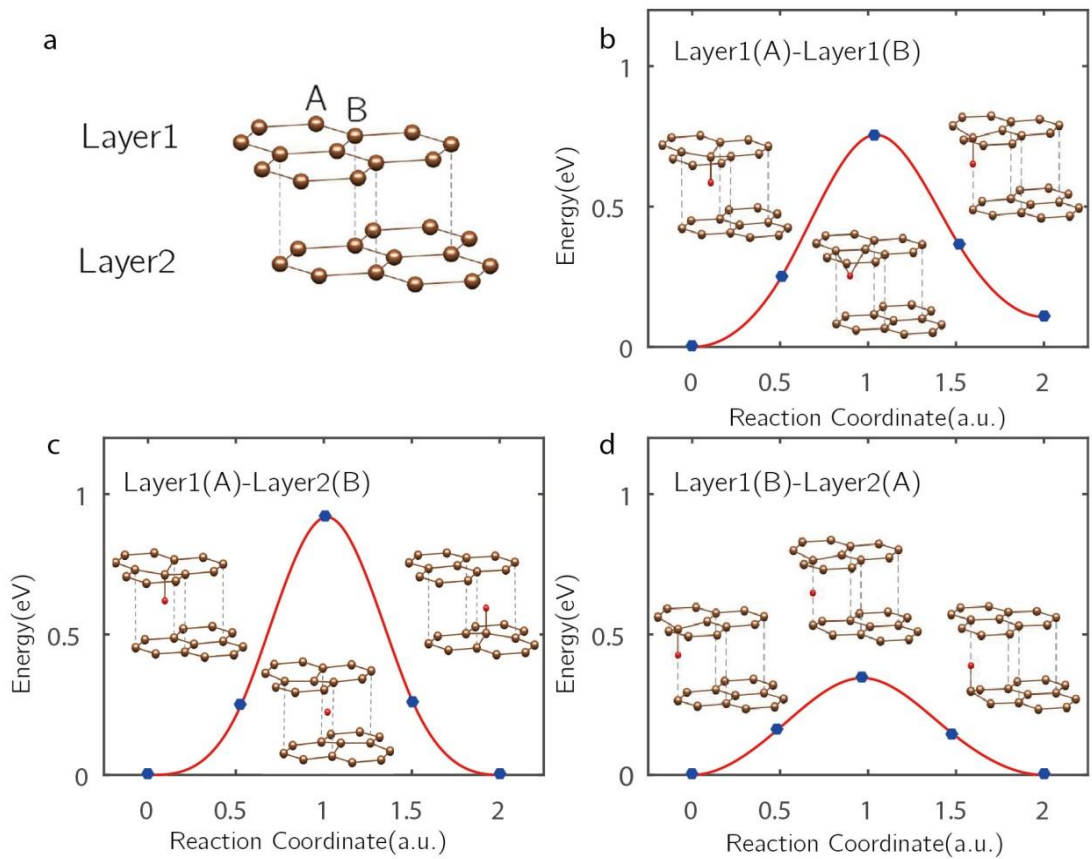

**Figure 5 | Energy barriers of diffusion and desorption for the confined H atom. (a)** Lattice structure of pristine bilayer graphene in Bernal stacking (side view). Two hexagonal sublattice are labeled as A and B. The red ball denotes H atom. **(b)** Computed barrier for migration of hydrogen atom from A site to B site on the top layer. **(c)** Hydrogen atom migration barrier from Layer1(A) sublattice to Layer 2(B) site. **(d)** Hydrogen atom migration barrier from Layer 1(B) site to Layer 2(A) site.

17